\documentclass[12pt]{article}
\usepackage{amsmath}
\usepackage[dvips]{graphicx}
\usepackage{epsfig}
\usepackage{amsmath}
\usepackage{amssymb}
\usepackage{graphics,epstopdf}
\usepackage{amsfonts}
\usepackage{amsthm}
\usepackage[usenames]{color}

\definecolor{AV}{rgb}{0.65,0.0,0}
\definecolor{GC}{rgb}{0,0.0,0.65}
\definecolor{WS}{rgb}{0,0.65,0}

\usepackage[
      colorlinks=true,
      linkcolor=blue,
      urlcolor=blue,
      filecolor=blue,
      citecolor=red,
      pdfstartview=FitV,
      pdftitle={},
      pdfauthor={},
      pdfsubject={},
      pdfkeywords={},
      pdfpagemode=None,
      bookmarksopen=true
]{hyperref}

\newcommand{\bm}{\begin{multiline}}
\newcommand{\beq}{\begin{equation}}
\newcommand{\eeq}{\end{equation}}
\newcommand{\beqs}{\begin{eqnarray}}
\newcommand{\eeqs}{\end{eqnarray}}

\newcommand{\ra}{\rightarrow}

\begin{document}

\thispagestyle{empty}

\hfill{}

\hfill{}

\hfill{}

\vspace{32pt}

\begin{center}

\textbf{\Large On double-black hole solutions of the Einstein-Maxwell-Dilaton theory in five dimensions}

\vspace{48pt}

\textbf{Cristian Stelea}\footnote{E-mail: \texttt{cristian.stelea@outlook.com}}

\vspace*{0.2cm}

\textit{Research Department, Faculty of Physics, ``Alexandru Ioan Cuza" University}\\[0pt]
\textit{11 Blvd. Carol I, Iasi, 700506, Romania}\\[.5em]

\end{center}

\vspace{30pt}

\begin{abstract}
We describe a solution-generating technique that will map a static charged solution of the Einstein-Maxwell theory in four (or five) dimensions to a five-dimensional solution of the Einstein-Maxwell-Dilaton theory. As examples of this technique first we show first how to construct the dilatonic version of the Reissner-Nordstr\"om solution in five dimensions and then we consider the more general case of the double black hole solutions and describe some of their properties. We found that in the general case the value of the conical singularities in between the black holes is affected by the dilaton's coupling constant to the gauge field and only in the particular case when all charges are proportional to the masses this dependence cancels out.
  
\end{abstract}

\vspace{32pt}

\setcounter{footnote}{0}

\newpage

\section{Introduction}

One of the most important predictions of General Relativity is the existence of black holes. Generally speaking, these are regions of spacetime where gravity is so strong that nothing can escape its grasp once it crossed the black hole event horizon. Black holes can be formed by gravitational collapse of massive stars and there is by now compelling evidence that such objects exist in the Universe (see for instance \cite{Genzel} - \cite{Abbott:2017vtc}  and also \cite{Yagi:2016jml}, \cite{Cardoso:2017njb} and references therein). One of the most interesting property of the black hole physics in four dimensions is the so-called `no hair' theorem (for a review see \cite{Chrusciel:2012jk}), which basically states that all regular asymptotically flat solutions of the Einstein-Maxwell (EM) equations are uniquely determined by their conserved asymptotic charges (such as mass, angular momentum and electric (magnetic) charges) and, moreover, they are included in the Kerr-Newman class of solutions. 

Recently, it was realized that higher than four dimensions black holes exhibit a much richer structure than their four-dimensional counterparts (see for reviews \cite{Emparan:2008eg},\cite{Obers:2008pj}). For instance, while in four dimensions a black hole can have only spherical topology of the event horizon \cite{Hawking:1971vc}, in higher dimensions black holes can have nontrivial horizon topologies. The most important example was provided in five dimensions by the black ring solution discovered by Emparan and Reall \cite{Emparan:2001wn} (for a review see \cite{Emparan:2006mm}), which has the $S^2\times S^1$ topology of the event horizon. Furthermore, the black ring solution can carry (in certain conditions) the same amount of mass and angular momenta as the five dimensional Myers-Perry spherical black hole \cite{Emparan:2004wy}. In consequence, in higher than four dimensions it is no longer true that stationary black holes are uniquely specified by their conserved charges at infinity. Using the inverse-scattering technique a  more general solution describing a rotating black ring, with rotation not only along $S^1$ but also along the azimuthal direction of $S^2$ of the ring horizon has been presented in \cite{Morisawa:2007di,Chen:2011jb} and demanding the absence of the conical singularities it reduces to the balanced rotating black ring found by Pomeransky and Senkov \cite{Pomeransky:2006bd}. This solution generalized the black rings found in \cite{Mishima:2005id,Figueras:2005zp}, which had rotation only along the azimuthal direction of $S^2$.  Following the discovery of the rotating black ring, its generalization to black Saturn \cite{Elvang:2007rd} and multi-black rings have been found in five dimensions \cite{Iguchi:2007is} - \cite{Izumi:2007qx}.  

Charged configurations of such objects have also been considered in literature: the black rings with electric charge has been studied in \cite{Kunduri:2004da}, while charged rings in string theory and supergravity have been considered in \cite{Elvang:2003yy} - \cite{Gauntlett:2004qy}.\footnote{Other general charged solutions in presence of dilaton fields were studied in \cite{Zangeneh:2015wia} - \cite{Sheykhi:2016tma}.} While the construction of vacuum multi-black hole solutions can be accomplished by using the inverse-scattering technique, solutions describing general charged multi-black hole configurations are more difficult to derive. This happens since the known solution generating methods generically lead to charged solutions having all charges proportional to the masses and, therefore, they cannot describe the most general solutions for which the individual charges and masses are independent parameters. Recently, a new solution generating technique has been proposed in \cite{Chng:2008sr} and it was used to construct asymptotically flat solutions describing a general double-Reissner-Nordstr\"om solution, a charged black Saturn as well as the charged double ring solutions \cite{Stelea:2011jm}. In order to construct multi-black hole configurations in spaces with Kaluza-Klein asymptotics it turns out that one has to modify this solution-generating technique as in \cite{Stelea:2009ur}. The double-Reissner-Nordstr\"om solution in this background has been constructed in \cite{Stelea:2009ur}, while more general solutions with Kaluza-Klein asymptotics have been derived in \cite{Stelea:2011fj}. Finally, solutions describing charged black holes on the Taub-bolt instanton have been studied in \cite{Stelea:2012ph}, \cite{Stelea:2012xg}, while charged multi-black holes on Kaluza-Klein bubbles have been derived in \cite{Kunz:2013osa}. 

One should note that even if the final solutions obtained using the solution-generating techniques in \cite{Chng:2008sr} and \cite{Stelea:2009ur} are solutions of the full Einstein-Maxwell-Dilaton field equations, with arbitrary coupling of the dilaton to the Maxwell field, in the above mentioned papers only the Einstein-Maxwell solutions have been discussed. However, when one considers the full solutions of the EMD system, that is when one turns on the dilaton field and its coupling to the Maxwell field, their interpretation as describing charged multi-black hole solutions is not the correct one unless one considers the extremally charged cases. The basic reason is that for a black hole horizon the $g_{tt}$ metric coefficient should have a single zero at the location of a black hole horizon, while in presence of the dilaton field, when the coupling constant to the Maxwell field is nonzero, the horizon location is a multiple root\footnote{We thank Gavin Hartnett for pointing this out.} and therefore it cannot describe the location of a black hole horizon except for the extremal case.

In this paper we try to remedy this situation by presenting a modification of the above solution-generating method, which will allow us to derive the general static charged multi-black hole solutions in asymptotically flat backgrounds, in the five-dimensional Einstein-Maxwell-Dilaton theory, with arbitrary coupling of the dilaton to the Maxwell field. 

The structure of our paper is as follows: in the next section we present the solution generating technique that will connect a static solution of the Einstein-Maxwell theory to a solution of the Einstein-Maxwell-Dilaton theory in five dimensions, with arbitrary coupling of the dilaton to the Maxwell field. In Section $2.1$ we present the generation of the dilatonic Reissner-Nordstr\"om solution in five dimensions as an example of this technique. In Section $3$ we construct the proper dilatonic generalization of the double black hole system, such as the double Reissner-Nordstr\"om solution and the dilatonic black Saturn and describe some of their properties. The final section is dedicated to conclusions and avenues for further work. 

\section{The solution generating technique in five dimensions}\label{sec2}

In this section we develop a new solution-generating technique that will map a general static solution of the Einstein-Maxwell equations in five dimensions to a five-dimensional solution of the EMD theory with arbitrary coupling of the dilaton to the electromagnetic field. To this end, we start with the five-dimensional Lagrangean of the EMD theory, which describes gravity coupled to a dilaton field $\phi$ and a $2$-form field strength $F_{(2)}$:
\beqs
{\cal L}_{(5)EMD}&=&\sqrt{-g}\bigg[R-\frac{1}{2}(\partial\phi)^2-\frac{1}{4} e^{a\phi}F_{(2)}^2\bigg],
\eeqs
where $F_{(2)}=dA_{(1)}$, while $A_{(1)}=\omega dt$ is the electromagnetic potential. At this point we assume that the metric and the matter fields are time-independent and the metric belongs to the generalized Weyl class of metrics \cite{Emparan:2001wk}. Let us perform now a Kaluza-Klein reduction along the time direction using the metric ansatz:
\beqs
ds_{(5)EMD}^2&=&-e^{-\frac{2\phi_1}{\sqrt{3}}}dt^2+e^{\frac{\phi_1}{\sqrt{3}}}ds_{(4)EMD}^2.
\eeqs
The metric $ds_{(4)EMD}^2$ is then a solution of the equations of motion derived from the dimensionally reduced Lagrangean:
\beqs
\label{L4EMD}
 {\cal L}_{(4)EMD}&=&\sqrt{-g_{(4)}}\bigg[R-\frac{1}{2}(\partial\phi_1)^2-\frac{1}{2}(\partial\phi)^2+\frac{1}{2}e^{a\phi+\frac{2\phi_1}{\sqrt{3}}}(\partial\omega)^2\bigg].
\eeqs
Consider now a further dimensional reduction along the $\chi$-coordinate using the metric ansatz:
\beqs
ds_{(4)EMD}&=&e^{-\phi_2}d\chi^2+e^{\phi_2}ds_{(3)EMD}^2,
\eeqs
which is a solution of the equations of motion described the dimensionaly reduced Lagrangean:
\beqs
\label{L3EMD}
{\cal L}_{(3)EMD}&=&\sqrt{-g_{(3)}}\bigg[R-\frac{1}{2}(\partial\phi_1)^2-\frac{1}{2}(\partial\phi_2)^2-\frac{1}{2}(\partial\phi)^2+\frac{1}{2}e^{a\phi+\frac{2\phi_1}{\sqrt{3}}}(\partial\omega)^2\bigg].
\eeqs
Let us perform now the following field redefinitions:
\beqs
\label{redefEMD}
\psi_1&=&\phi_1+\frac{a\sqrt{3}}{2}\phi,~~~~\psi_2=-\frac{a\sqrt{3}}{2}\phi_1+\phi,~~~~\psi_3=\sqrt{\frac{4+3a^2}{4}}\phi_2,~~~\omega_1=\sqrt{\frac{4+3a^2}{4}}\omega.
\eeqs
Then the matter-only part of the Lagrangean (\ref{L3EMD}) becomes:
\beqs
\label{matEMD}
{\cal L}_{EMD}^{matter}&=&\frac{4}{4+3a^2}\bigg[-\frac{1}{2}(\partial\psi_1)^2-\frac{1}{2}(\partial\psi_2)^2-\frac{1}{2}(\partial\psi_3)^2+\frac{1}{2}e^{\frac{2\psi_1}{\sqrt{3}}}(\partial\omega_1)^2\bigg].
\eeqs

Consider now the general static solution of the Einstein-Maxwell theory in five dimensions found previously in \cite{Chng:2008sr}, which was derived starting from a similar four-dimensional solution of the Einstein-Maxwell theory. In the most general form, it can be written as:\footnote{We refer the reader to \cite{Chng:2008sr} for more details.}
\begin{eqnarray}  \label{rel1}
ds_{5}^2&=&-fdt^2+f^{-\frac{1}{2}}\bigg[e^{2h}d\chi^2+e^{-2h}\big[e^{3\lambda/2+2\gamma}(d\rho^2+dz^2)+\rho^2d\varphi^2\big]\bigg],
\notag \\
A_t&=&\omega_0.  
\end{eqnarray}
Here $h$ is an arbitrary harmonic function\footnote{That is, it satisfies the equation $\nabla^2h=\frac{\partial^2h}{\partial\rho^2}+\frac{1}{\rho}\frac{\partial h}{\partial\rho}+\frac{\partial^2h}{\partial z^2}=0.$}, while its backreaction in the metric is taken care by means of the function $\gamma$, which satisfies:
 \begin{eqnarray}  \label{gammap1a}
\partial_\rho{\gamma}&=&\rho[(\partial_\rho h)^2-(\partial_z h)^2],~~~~~~~
\partial_z{\gamma}=2\rho(\partial_\rho h)(\partial_z h).
\end{eqnarray}
Recall now that the solution (\ref{rel1}) was derived in \cite{Chng:2008sr} starting from a seed-solution of the Einstein-Maxwell theory in four dimensions, which had the following form:
 \begin{eqnarray}  \label{metric4da}
ds_{4}^{2} &=&-fdt^{2}+f^{-1}\big[e^{2\lambda}(d\rho
^{2}+dz^{2})+\rho ^{2}d\varphi ^{2}\big],  \notag \\
{A_{(1)}} &=&\frac{2}{\sqrt{3}}\omega_0 dt.
\end{eqnarray} 
Performing now two dimensional reductions, first along the timelike direction $t$, then along the $\chi$ coordinate, one obtains the following three-dimensional metric
\beqs
\label{me3EM}
ds_{(3)EM}^2&=&e^{\frac{3\lambda}{2}+2\gamma}(d\rho^2+dz^2)+\rho^2d\varphi^2,
\eeqs
which is a solution of the field equations derived from the following dimensionally reduced Lagrangean:
\beqs
\label{matEM}
{\cal L}_{(3)EM}^{matter}&=&-\frac{1}{2}(\partial\varphi_1)^2-\frac{1}{2}(\partial\varphi_2)^2+\frac{1}{2}e^{\frac{2\varphi_1}{\sqrt{3}}}(\partial\omega_0)^2,
\eeqs
where we defined the fields:
\beqs
e^{-\frac{\varphi_1}{\sqrt{3}}}&=&f,~~~~~~\varphi_2=-2h.
\eeqs
Since the matter Lagrangean of the EMD theory has three scalar fields, one should simply add a new scalar field $\varphi_3$ to (\ref{matEM})\footnote{Note that $\varphi_3$ is a harmonic function, according to its equations of motion derived from (\ref{matEMf}).} 
\beqs
\label{matEMf}
{\cal L}_{(3)EM}^{matter}&=&-\frac{1}{2}(\partial\varphi_1)^2-\frac{1}{2}(\partial\varphi_2)^2-\frac{1}{2}(\partial\varphi_3)^2+\frac{1}{2}e^{\frac{2\varphi_1}{\sqrt{3}}}(\partial\omega_0)^2
\eeqs
and also modify the metric (\ref{me3EM}) by including a function $\tau$ to accommodate its backreaction in the Einstein equations:
\beqs
\label{me3EMf}
ds_{(3)EM}^2&=&e^{\frac{3\lambda+\tau}{2}+2\gamma}(d\rho^2+dz^2)+\rho^2d\varphi^2.
\eeqs
Here the function $\nu$ satisfies the equations:
 \begin{eqnarray}  \label{nu1a}
\partial_\rho{\tau}&=&\rho[(\partial_\rho \varphi_3)^2-(\partial_z \varphi_3)^2],~~~~~~~
\partial_z{\tau}=2\rho(\partial_\rho \varphi_3)(\partial_z \varphi_3).
\end{eqnarray}  

Noting now that the matter-only parts of the two dimensionally reduced Lagrangeans of the two theories have the same functional form, up to a proportionality constant, in order to match the solutions of their equations of motion when coupled to gravity we have to further modify the three-dimensional geometry (\ref{me3EMf}) such that its Ricci tensor is also rescaled by the constant factor $\frac{4}{4+3a^2}$. In consequence, if one performs the following identifications:
\beqs
\psi_1&=&\varphi_1,~~~~\psi_2=\varphi_2,~~~~\psi_3=\varphi_3,~~~~\omega_1=\omega_0
\eeqs
then the three-dimensional metric:
\beqs
ds_{(3)EMD}^2&=&\left(e^{\frac{3\lambda+\tau}{2}+2\gamma}\right)^{\frac{4}{4+3a^2}}(d\rho^2+dz^2)+\rho^2d\varphi^2
\eeqs
will provide us with a solution of the equations of motion derived from the Lagrangean (\ref{matEMD}). We have now all the necessary ingredients to reconstruct the final five-dimensional solution of the EMD theory. 

To summarize our results: starting with the general solution given in (\ref{rel1}) (or the seed (\ref{metric4da})), then the final EMD solution can be written as:
\beqs
\label{finalsolRN}
ds^2&=&-\frac{f^{\frac{4}{3a^2+4}}}{\left(e^{2h}\right)^{\frac{4a}{3a^2+4}}}dt^2+f^{-\frac{2}{3a^2+4}}\left(e^{2h}\right)^{\frac{2a}{3a^2+4}}\bigg[e^{2H}d\chi^2+\nonumber\\
&&+e^{-2H}\bigg[e^{\frac{6\lambda}{4+3a^2}+2\tau+2\gamma\frac{4}{4+3a^2}}(d\rho^2+dz^2)+\rho^2d\varphi^2\bigg]\bigg],\nonumber\\
e^{\phi}&=&f^{-\frac{3a}{3a^2+4}}\left(e^{-2h}\right)^{\frac{4}{3a^2+4}},~~~~A_{(1)}=\omega dt,
\eeqs
where:
\beqs
\omega=\sqrt{\frac{4}{3a^2+4}}\omega_0
\eeqs

In this final solution the functions $h$ and $H$ are arbitrary harmonic functions, while the functions $\gamma$ and $\tau$ satisfy respectively the equations (\ref{gammap1a}) and:
 \begin{eqnarray}  \label{tau1a}
\partial_\rho{\tau}&=&\rho[(\partial_\rho H)^2-(\partial_z H)^2],~~~~~~~
\partial_z{\tau}=2\rho(\partial_\rho H)(\partial_z H).
\end{eqnarray} 

In writing the final solution, for further convenience we have redefined the harmonic function $\varphi_3=-\sqrt{3a^2+4}H$ and rescaled the function $\tau$ in order to satisfy the relations (\ref{tau1a}).

Note that we have obtained a more general solution than the one previously derived in \cite{Chng:2008sr}. As a check of our solution generating technique, if one takes the coupling constant to zero $a=0$, then the dilaton field $\phi$ decouples from the Maxwell field and its backreaction in the metric is taken into account by means of the function $\mu$. In order to recover the initial solution of the Einstein-Maxwell theory one has to take the arbitrary harmonic function $h=0$, which leads to $\mu=0$ and one obtains the solution (\ref{rel1}), as expected.

\subsection{The dilatonic Reissner-Nordstr\"om solution in five dimensions}

Our starting point will be the four-dimensional Reissner-Nordstr\"{o}m solution, which is written in Weyl form
as \cite{Emparan:2001bb}:
\begin{eqnarray}  \label{RN4dim}
ds^2&=&-fdt^2+f^{-1} \big[e^{2\lambda}(d\rho^2+dz^2)+\rho^2 d\varphi^2\big], \\
\omega_0&=&-\frac{4q}{r_1+r_2+2m},~~~~~~~f=\frac{(r_{1}+r_{2})^2-4\sigma^2}{(r_{1}+r_{2}+2m)^2},~~~~~~~ e^{2\lambda}=\frac{(r_{1}+r_{2})^2-4\sigma^2}{4r_{1}r_{2}},  \notag
\end{eqnarray}
where
\begin{eqnarray}
r_{1}&=&\sqrt{\rho^2+(z-\sigma)^2}, ~~~~~~~ r_{2}=\sqrt{\rho^2+(z+\sigma)^2}.
\end{eqnarray}
Note that $\sigma=\sqrt{m^2-q^2}$, while $m$ denotes the mass and $q$ is the electric charge.

Applying the solution-generating technique from the previous section one arrives at (\ref{finalsolRN}) where now:
\beqs
\omega&=&-\sqrt{\frac{4}{3a^2+4}}\frac{2\sqrt{3}q}{r_1+r_2+2m}.
\eeqs
So far we have kept the harmonic functions $h$ and $H$ arbitrary. In order to construct the dilatonic Reissner-Nordstr\"om solution one has to make some educated guess regarding their form. Note that $H$ can alter the metric only along the spacelike directions and its form can be carefully chosen in order to impose that the background geometry becomes asymptotically flat. Since its form is not affected by the presence of the dilaton field, we shall use the harmonic function used to construct the five-dimensional Reissner-Nordstr\"om solution in the EM theory \cite{Chng:2008sr}:
\beqs
e^{2H}&=&(r_{2}+(z+\sigma))\left(\frac{r_{1}+(z-\sigma)} {r_{2}+(z+\sigma)}\right)^{\frac{1}{2}}\cr &=&\left[(r_{2}+\zeta_2)(r_{1}+\zeta_1)\right]^{\frac{1}{2}},
\end{eqnarray}
and we can now find $\tau$ from (\ref{nu1a}):\footnote{Note that $e^{2\tau}$  and $e^{\mu}$ are defined up to a multiplicative constant and in the followings we have chosen these constants for further convenience.}
\begin{eqnarray}
e^{2\tau}&=&\frac{[(r_{2}+\zeta_2)(r_{1}+\zeta_1)]^{\frac{1}{2}}} {[8r_{1}r_{2}Y_{12}]^{\frac{1}{4}}}.
\end{eqnarray}
where $\zeta_1=z-\sigma$, $\zeta_2=z+\sigma$ and $Y_{12}=r_1r_2+\zeta_1\zeta_2+\rho^2$. The first factor in $e^{2h}$ moves the semi-infinite rod $z<-\sigma$ from the $\varphi$ direction to the $\chi$ direction, while the second factor corresponds to a `correction' of the black hole horizon. Once we identified the black hole `correction' factor we can finally guess the form of the remaining harmonic function $h$:
\beqs
e^{2h}&=&\left(\frac{r_{1}+(z-\sigma)} {r_{2}+(z+\sigma)}\right)^{-\frac{3a}{4}}
\eeqs 
and integrate (\ref{gammap1a}) to obtain:
\beqs
e^{2\gamma}&=&\left(\frac{Y_{12}}{2r_1r_2}\right)^{\frac{9a^2}{16}}\equiv\left(e^{2\lambda}\right)^{\frac{9a^2}{16}}.
\eeqs
Also, one should note the identity $2Y_{12}=(r_1+r_2)^2-4\sigma^2$. We are now ready to re-assemble the final solution in five dimensions. To simplify the computations, let us first recast the general solution in the following form:
\beqs
\label{finalsolRNnew}
ds^2&=&-\frac{f}{F^{2a}}dt^2+F^{a} f^{-\frac{1}{2}}\bigg[e^{2H}d\chi^2+e^{-2H}\bigg[e^{\frac{3\lambda}{2}+2\tau}(d\rho^2+dz^2)+\rho^2d\varphi^2\bigg]\bigg],\nonumber\\
e^{\phi}&=&\frac{1}{F^2},~~~~\omega=-\sqrt{\frac{4}{3a^2+4}}\frac{2\sqrt{3}q}{r_1+r_2+2m},
\eeqs
where we defined:
\beqs
F&=&\left(f\frac{r_2+\zeta_2}{r_1+\zeta_1}\right)^{\frac{3a}{3a^2+4}}.
\eeqs
Finally, in order to show that the generated solution is indeed the dilatonic Reissner-Nordstr\"om solution in five dimensions we have to convert it from cylindrical coordinates $(\rho, z)$ to polar coordinates $(r,\theta)$ by using the relations \cite{Emparan:2001wk}:
\begin{eqnarray}
\rho^2&=&r^2(r^2-4\sigma)\sin^2\theta\cos^2\theta,~~~~~~~ z=\frac{1}{2}%
(r^2-2\sigma)\cos2\theta.  \label{coordtransfsingleRN}
\end{eqnarray}
We obtain:
\begin{eqnarray}
ds_{5}^2&=&-\frac{r^2(r^2-4\sigma)}{(r^2+2(m-\sigma))^2}\frac{dt^2}{F^{2a}} +F^a\frac{
r^2+2(m-\sigma)}{r^2}\Big(\frac{r^2}{r^2-4\sigma}dr^2+r^2(d\theta^2+\sin^2%
\theta d\varphi^2+\cos^2\theta d\chi^2)\Big)  \notag \\
&&  \notag \\
e^{\phi}&=&F^{-2},~~~~~~\omega=-\sqrt{\frac{4}{3a^2+4}}\frac{2\sqrt{3}\sqrt{m^2-\sigma^2}}{r^2+2(m-\sigma)},
\end{eqnarray}
where
\begin{eqnarray}
F&=&\left(\frac{r_1+r_2+2\sigma}{r_1+r_2+2m}\right)^{\frac{6a}{3a^2+4}}=\left(\frac{r^2}{r^2+2(m-\sigma)}\right)^{\frac{6a}{3a^2+4}},
\end{eqnarray}
which is indeed the five-dimensional dilatonic Reissner-Nordstr\"{o}m solution of the EMD theory with arbitrary coupling of the dilaton to the Maxwell field.

\section{The asymptotically flat dilatonic double-Reissner-Nordstr\"om solutions in five dimensions}

According to the solution-generating method from Section $2$, we shall start now from the four-dimensional double-Reissner-Nordstr\"om solution of the Einstein-Maxwell theory, in the parameterization given recently by Manko in \cite{Manko:2007hi}:
\beqs
\label{Manko4d}
ds^2&=&-fdt^2+f^{-1}[e^{2\lambda}(d\rho^2+dz^2)+\rho^2d\varphi^2], ~~~~A_{(1)}=\Psi dt,
\eeqs
Here
\begin{equation}
f=\frac{A^{2}-B^{2}+C^{2}}{(A+B)^{2}},~~~~~~e^{2\lambda}=\frac{
A^{2}-B^{2}+C^{2}}{16\sigma _{1}^{2}\sigma _{2}^{2}(\nu
+2k)^{2}r_{1}r_{2}r_{3}r_{4}},~~~~~~~\Psi=-\frac{2C}{A+B},  
\end{equation}%
where:
\begin{eqnarray}
A &=&\sigma _{1}\sigma _{2}[\nu
(r_{1}+r_{2})(r_{3}+r_{4})+4k(r_{1}r_{2}+r_{3}r_{4})]-(\mu ^{2}\nu
-2k^{2})(r_{1}-r_{2})(r_{3}-r_{4}),  \notag \\
B &=&2\sigma _{1}\sigma _{2}[(\nu M_{1}+2kM_{2})(r_{1}+r_{2})+(\nu
M_{2}+2kM_{1})(r_{3}+r_{4})]  \notag \\
&&-2\sigma _{1}[\nu \mu (Q_{2}+\mu )+2k(RM_{2}+\mu Q_{1}-\mu
^{2})](r_{1}-r_{2})  \notag \\
&&-2\sigma _{2}[\nu \mu (Q_{1}-\mu )-2k(RM_{1}-\mu Q_{2}-\mu
^{2})](r_{3}-r_{4}),  \notag \\
C &=&2\sigma _{1}\sigma _{2}\{[\nu (Q_{1}-\mu )+2k(Q_{2}+\mu
)](r_{1}+r_{2})+[\nu (Q_{2}+\mu )+2k(Q_{1}-\mu )](r_{3}+r_{4})\}  \notag \\
&&-2\sigma _{1}[\mu \nu M_{2}+2k(\mu M_{1}+RQ_{2}+\mu R)](r_{1}-r_{2})
\notag \\
&&-2\sigma _{2}[\mu \nu M_{1}+2k(\mu M_{2}-RQ_{1}+\mu R)](r_{3}-r_{4}),
\end{eqnarray}%
with constants:
\begin{eqnarray}
\nu &=&R^{2}-\sigma _{1}^{2}-\sigma _{2}^{2}+2\mu
^{2},~~~~~~~k=M_{1}M_{2}-(Q_{1}-\mu )(Q_{2}+\mu ),  \notag \\
\sigma _{1}^{2} &=&M_{1}^{2}-Q_{1}^{2}+2\mu Q_{1},~~~~~~~\sigma
_{2}^{2}=M_{2}^{2}-Q_{2}^{2}-2\mu Q_{2},~~~~~~~\mu =\frac{%
M_{2}Q_{1}-M_{1}Q_{2}}{M_{1}+M_{2}+R},
\end{eqnarray}%
while $r_{i}=\sqrt{\rho ^{2}+\zeta _{i}^{2}}$, for $i=1..4$, with:
\begin{equation}
\zeta _{1}=z-\frac{R}{2}-\sigma _{2},~~~~~\zeta _{2}=z-\frac{R}{2}+\sigma
_{2},~~~~~\zeta _{3}=z+\frac{R}{2}-\sigma _{1},~~~~~\zeta _{4}=z+\frac{R}{2}%
+\sigma _{1}.
\label{ai}
\end{equation}%
This solution describes the superposition of two Reissner-Nordstr\"om black holes in four dimensions and it is parameterized by five independent parameters, which correspond to the physical values of the electric charges $Q_{1,2}$, masses $M_{1,2}$ of the two black holes, while $R$ describes the coordinate distance separating the two black hole horizons on the axis. For a more detailed discussion of the properties of this solution we refer the interested reader to \cite{Manko:2008gb}. In general, the function $e^{2\lambda}$ can be determined up to a
constant and its precise numerical value has been fixed here by allowing the presence of conical singularities only in the portion in between the black holes along the $\varphi $ axis. Consequently one has:
\begin{equation}
e^{2\lambda}|_{\rho =0}=\left( \frac{\nu -2k}{\nu +2k}\right) ^{2},
\label{strutManko}
\end{equation}%
for $-R/2+\sigma _{1}<z<R/2-\sigma _{2}$ and $e^{2\lambda}|_{\rho =0}=1$ elsewhere.

Before we apply the solution generating technique from Section \ref{sec2}, for further convenience, let us perform first a rescalling of the harmonic function of the form $h\ra-\frac{3a}{2}h$, which amounts to a rescalling of $\gamma\ra\frac{9a^2}{4}\gamma$, such that the final EMD solution can be written in the following form:
\beqs
\label{finalsolEMDa}
ds^2&=&-\frac{f^{\frac{4}{3a^2+4}}}{\left(e^{2h}\right)^{-\frac{6a^2}{3a^2+4}}}dt^2+f^{-\frac{2}{3a^2+4}}\left(e^{2h}\right)^{-\frac{3a^2}{3a^2+4}}\bigg[e^{2H}d\chi^2+\nonumber\\
&&+e^{-2H}\bigg[e^{\frac{6\lambda}{3a^2+4}+2\tau+2\gamma\frac{9a^2}{3a^2+4}}(d\rho^2+dz^2)+\rho^2d\varphi^2\bigg]\bigg],\nonumber\\
e^{\phi}&=&f^{-\frac{3a}{3a^2+4}}\left(e^{2h}\right)^{-\frac{6a}{3a^2+4}},~~~~A_{(1)}=\omega dt,
\eeqs
where
\beqs
\omega&=&-\sqrt{\frac{12}{3a^2+4}}\frac{C}{A+B}.
\eeqs

It is now obvious that if one takes the coupling constant $a=0$ then one recovers the EM solution discussed in \cite{Chng:2008sr}. As in the dilatonic single black hole from the previous section, we have to chose carefully the form of the two harmonic functions $h$ and $H$ in order to ensure that the final solution describes a system of two dilatonic black holes in five dimensions. We shall discuss  two specific choices for such functions: the first will describe the dilatonic generalization of two spherical black holes, while the second choice will describe the dilatonic generalization of the black saturn solution from \cite{Chng:2008sr}, that is of a spherical black hole surrounded by a black ring.

\subsection{The dilatonic double-Reissner-Nordstr\"{o}m solution}

Based on the experience with the single spherical black hole, we shall pick the following form of the harmonic function $H$:
\begin{eqnarray}
e^{2H}&=&\frac{\sqrt{(r_1+\zeta_1)(r_2+\zeta_2)(r_3+\zeta_3)(r_4+\zeta_4)}}{%
r_0+\zeta_0},
\end{eqnarray}
where we denote $r_i=\sqrt{\rho^2+\zeta_i^2}$ and $\zeta_i=z-a_i$ for $i=0 ..
4$, $a_i$ can be read from (\ref{ai}) and $a_0=0$. With this choice one can integrate (\ref{tau1a}) to obtain:
\begin{eqnarray}
e^{2\tau-2H}&=&\frac{1}{K_0r_0}\frac{\left(Y_{01}Y_{02}Y_{03}Y_{04}%
\right)^{\frac{1}{2}}}{%
\left(r_1r_2r_3r_4Y_{12}Y_{13}Y_{14}Y_{23}Y_{24}Y_{34}\right)^{\frac{1}{4}}}
\end{eqnarray}
Here $K_0$ is an arbitrary constant whose value will be fixed later on, and $Y_{ij}=r_ir_j+\zeta_i\zeta_j+\rho^2$. Taking into account the `correction' factor for each black hole horizon one can finally pick the second harmonic function to be of the form:
\begin{eqnarray}
\label{hcorrect}
e^{2h}&=&\sqrt{\frac{(r_1+\zeta_1)(r_3+\zeta_3)}{(r_2+\zeta_2)(r_4+\zeta_4)}}.
\end{eqnarray}
One easily integrates (\ref{gammap1a}) to find:
\begin{eqnarray}
e^{2\gamma}&=&\left(\frac{Y_{12}Y_{14}Y_{23}Y_{34}}{4
r_1r_2r_3r_4Y_{13}Y_{24}}\right)^{\frac{1}{4}},
\end{eqnarray}
Let us consider now the rod structure of this solution. Following the procedure from \cite{Chen:2010zu}, let us note that one has five turning points such that $a_i$, $i=0..4$ coincide with the turning points of the seed solution (\ref{Manko4d}), while the fifth is $a_0=0$. They divide the z-axis into six rods, which can be described in the following way, specifying the rod direction vectors  with respect to a basis of Killing fields $\{\frac{\partial}{\partial t}, \frac{\partial}{\partial \phi}, \frac{\partial}{\partial\chi}\}$, and  normalizing them to the surface gravity of each fixed point set:
\begin{itemize}
\item \textbf{Rod 1} - For $z<a_4$ one has a semi-infinite spacelike rod with normalized direction
\beqs
l_1=\sqrt{\frac{2\sqrt{2}}{K_0}}(0,0,1).
\eeqs
\item \textbf{Rod 2} - For $a_4<z<a_3$ one has a finite timelike rod that corresponds to a black hole horizon  with $S^3$ topology. Its normalized rod direction is given by
$l_2=\frac{1}{k_{H}^1}(1,0,0)$, where
\beqs
k_{H}^1=\sqrt{\frac{K_0}{2\sqrt{2}}}p_1\bigg[\left(\rho f^{-1}e^{\lambda}\right)|_{H_1}\bigg]^{-\frac{3}{3a^2+4}}P_1^{\frac{9a^2}{3a^2+4}}
\eeqs
 is the surface gravity on the black hole horizon represented by this rod.
\item \textbf{Rod 3} - For $a_3<z<a_0$ one has a finite spacelike rod with normalized direction
$l_3=\frac{1}{k_{1}}(0,1,0)$,
where
\beqs
k_{1}=\frac{1}{2}\sqrt{\frac{K_0}{2\sqrt{2}}}\left|\frac{\nu+2k}{\nu-2k}\right|^{{\frac{3}{3a^2+4}}}
\left(\frac{(R^2-4\sigma_1^2)^2}{((R+\sigma_2)^2-\sigma_1^2)((R-\sigma_2)^2-\sigma_1^2)}\right)^{\frac{1}{4}}\left(\frac{R^2-(\sigma_1-\sigma_2)^2}{R^2-(\sigma_1+\sigma_2)^2}\right)^{\frac{9a^2}{4(3a^2+4)}}
\eeqs
\item \textbf{Rod 4} - For $a_0<z<a_2$ one has again a finite spacelike rod, this time with normalized direction $l_4=\frac{1}{k_{2}}(0,0,1)$,
where
\beqs
k_{2}=\frac{1}{2}\sqrt{\frac{K_0}{2\sqrt{2}}}\left|\frac{\nu+2k}{\nu-2k}\right|^{{\frac{3}{3a^2+4}}}
\left(\frac{(R^2-4\sigma_2^2)^2}{((R+\sigma_1)^2-\sigma_2^2)((R-\sigma_1)^2-\sigma_2^2)}\right)^{\frac{1}{4}}\left(\frac{R^2-(\sigma_1-\sigma_2)^2}{R^2-(\sigma_1+\sigma_2)^2}\right)^{\frac{9a^2}{4(3a^2+4)}}
\eeqs

\item \textbf{Rod 5} - For $a_2<z<a_1$ one has a finite timelike rod, corresponding to a second black hole horizon with $S^3$ topology. Its normalized rod direction is found to be
$l_4=\frac{1}{k_{H}^2}(1,0,0)$, where
\beqs
k_{H}^2=\sqrt{\frac{K_0}{2\sqrt{2}}}p_2\bigg[\left(\rho f^{-1}e^{\lambda}\right)|_{H_2}\bigg]^{-\frac{3}{3a^2+4}}P_2^{\frac{9a^2}{3a^2+4}}
\eeqs
is the surface gravity of the black hole horizon corresponding to this rod.
\item \textbf{Rod 6} - for $z>a_1$ one has a semi-infinite spacelike rod with normalized direction
\beqs
l_6=\sqrt{\frac{2\sqrt{2}}{K_0}}(0,1,0).
\eeqs
\end{itemize}

Here we defined the following quantities:
\beqs
p_1&=&\left(\frac{4\sigma_1\big[(R+\sigma_1)^2-\sigma_2^2\big]}{(R+2\sigma_1)^2}\right)^{\frac{1}{4}},~~~~~p_2=\left(\frac{4\sigma_2\big[(R+\sigma_2)^2-\sigma_1^2\big]}{(R+2\sigma_2)^2}\right)^{\frac{1}{4}},
\label{pmic}
\eeqs
which appear in the following expansions on each black hole horizon:
\beqs
\rho^{\frac{1}{2}}e^{2\tau-2H}&=&\frac{2\sqrt{2}}{K_0}\frac{1}{p_i^2}.
\eeqs
Similarly, the constants:
\beqs
P_1&=&\left(\frac{R+\sigma_1-\sigma_2}{4\sigma_1(R+\sigma_1+\sigma_2)}\right)^{\frac{1}{4}},~~~~~P_2=\left(\frac{R+\sigma_2-\sigma_1}{4\sigma_2(R+\sigma_1+\sigma_2)}\right)^{\frac{1}{4}},
\label{pmare}
\eeqs
appear when expanding on each black hole horizon the expressions:
\beqs
\label{pmaree}
\rho^{\frac{1}{2}}e^{2\gamma-2h}&=&\frac{1}{P_i^2}
\eeqs
Finally, for each black hole horizon one has \cite{Manko:2008gb}:
\beqs
\left(\rho f^{-1}e^{\lambda}\right)|_{H_1}&=&\frac{\big[(R+M_1+M_2)(M_1+\sigma_1)-Q_1(Q_1+Q_2)\big]^2}{\sigma_1[(R+\sigma_1)^2-\sigma_2^2]},\nonumber\\
\left(\rho f^{-1}e^{\lambda}\right)|_{H_2}&=&\frac{\big[(R+M_1+M_2)(M_2+\sigma_2)-Q_2(Q_1+Q_2)\big]^2}{\sigma_2[(R+\sigma_2)^2-\sigma_1^2]}.
\label{rhofmu}
\eeqs
It should be obvious now that the integration constant $K_0$ should be chosen equal to $2\sqrt{2}$ in order for the solution to be asymptotically flat. It is then clear that with these choices one obtains the proper dilatonic generalization of the double-Reissner-Nordstr\"{o}m solution in five dimensions. This solution is described by five dimensionfull parameters, which roughly correspond to the masses $M_{1,2}$ and the charges $Q_{1,2}$ for each black hole and the coordinate distance $R$ along the $z$-axis, between their horizons. Certain restrictions have to imposed on these parameters to ensure that conical singularities do not occur. In our case, it turns out that the system will be in equilibrium iff the conditions $k_1=1$ and $k_2=1$ are satisfied. The effect of the dilaton field is taken into account by the presence of the coupling constant $a$ in these expressions. Unlike the Einstein-Maxwell case considered in \cite{Chng:2008sr}, \textit{apriori} one could hope that one could tune in the value of this extra parameter to satisfy these conditions, however we have been unable to find physically meaningful values of the parameters for which this happens (even in the extremally charged cases). In particular, the physical conditions that we asked were that the masses of the two black holes are positive $M_i>0$ and also the condition that the two black hole horizons do not overlap $R>\sigma_1+\sigma_2$.

\subsection{The dilatonic Black Saturn}

In order to construct a Black Saturn system consisting of a black ring with a spherical black hole in its center one should pick the following harmonic function:
\begin{eqnarray}  \label{H-Saturn}
e^{2H}&=&\sqrt{\frac{(r_1+\zeta_1)(r_3+\zeta_3)(r_4+\zeta_4)}{(r_2+\zeta_2)}}.
\end{eqnarray}
One can easily integrate (\ref{tau1a}) to find:
\begin{eqnarray}  \label{2gh-Saturn}
e^{2\tau-2H}&=&\frac{1}{K_0}\left(\frac{Y_{12}Y_{23}Y_{24}}{r_1r_2r_3r_4Y_{13}Y_{14}Y_{34}}\right)^{\frac{1}{4}},
\end{eqnarray}
where $K_0$ is a constant to be fixed later. Since the harmonic function $h$ contains only the `corrections' associated with each black hole horizon, the appropriate choice turns out to be again that from (\ref{hcorrect}).

The rod structure of this solution can be constructed using the same procedure from the dilatonic double Reissner-Nordstr\"{o}m case. One has now four turning points, that divide the $z$-axis into five rods, as follows:
\begin{itemize}
\item \textbf{Rod 1} - For $z<a_4$ one has a semi-infinite spacelike rod with normalized direction
\beqs
l_1=\sqrt{\frac{2}{K_0}}(0,0,1).
\eeqs
\item \textbf{Rod 2} - For $a_4<z<a_3$ one has a finite timelike rod that corresponds to a black hole horizon  with $S^3$ topology. Its normalized rod direction is given by
$l_2=\frac{1}{k_{H}^1}(1,0,0)$, where
\beqs
k_{H}^1=\sqrt{\frac{K_0}{2}}p_1\bigg[\left(\rho f^{-1}e^{\lambda}\right)|_{H_1}\bigg]^{-\frac{3}{3a^2+4}}P_1^{\frac{9a^2}{3a^2+4}}
\eeqs
 is the surface gravity on the black hole horizon represented by this rod.
\item \textbf{Rod 3} - For $a_3<z<a_2$ one has a finite spacelike rod with normalized direction
$l_3=\frac{1}{k_{3}}(0,1,0)$,
where
\beqs
k_{3}=\sqrt{\frac{K_0}{2}}\left|\frac{\nu+2k}{\nu-2k}\right|^{{\frac{3}{3a^2+4}}}
\left(\frac{(R+\sigma_2)^2-\sigma_1^2}{(R-\sigma_2)^2-\sigma_1^2}\right)^{\frac{1}{4}}\left(\frac{R^2-(\sigma_1-\sigma_2)^2}{R^2-(\sigma_1+\sigma_2)^2}\right)^{\frac{9a^2}{4(3a^2+4)}}
\eeqs
\item \textbf{Rod 4} - For $a_2<z<a_1$ one has a finite timelike rod, corresponding to a black ring horizon, with $S^1\times S^2$ topology. Its normalized rod direction is found to be
$l_4=\frac{1}{k_{H}^2}(1,0,0)$, where
\beqs
k_{H}^2=\sqrt{\frac{K_0}{2}}p_2\bigg[\left(\rho f^{-1}e^{\lambda}\right)|_{H_2}\bigg]^{-\frac{3}{3a^2+4}}P_2^{\frac{9a^2}{3a^2+4}}
\eeqs
is the surface gravity of the black hole horizon corresponding to this rod.
\item \textbf{Rod 5} - for $z>a_1$ one has a semi-infinite spacelike rod with normalized direction
\beqs
l_5=\sqrt{\frac{2}{K_0}}(0,1,0).
\eeqs
\end{itemize}
Here the constants $p_i$ are defined by:
\beqs
p_1&=&\left(\frac{4\sigma_1(R+\sigma_1+\sigma_2)}{(R+\sigma_1-\sigma_2}\right)^{\frac{1}{4}},~~~~~p_2=\left(\frac{(R+\sigma_2)^2-\sigma_1^2}{\sigma_2}\right)^{\frac{1}{4}},
\label{pmicsaturn}
\eeqs
which appear in the following expansions of the metric functions $e^{2\tau-2H}$ on each black hole horizon:
\beqs
\rho^{\frac{1}{2}}e^{2\tau-2H}&=&\frac{2}{K_0}\frac{1}{p_i^2}.
\eeqs
Similarly, the constants $P_i$ have the same values as in (\ref{pmare}) and they appear in the expansions (\ref{pmaree}) on each black hole horizon. In order for this solution to be asymptotically flat one has to pick the value of the constant $K_0=2$.

This solution is parameterized by five independent parameters that correspond physically to the masses and charges of the black hole that sits in the center of a black ring, while $R$ corresponds roughly to the ring radius. To ensure that conical singularities do not occur in this solution one has to impose certain restrictions on the vales of these parameters. For the dilatonic Black Saturn, there will be no conical singularities in between the black ring and the black hole if the condition $k_3=1$ is satisfied, that is:
\beqs
\left|\frac{\nu+2k}{\nu-2k}\right|^{{\frac{3}{3a^2+4}}}
\left(\frac{(R+\sigma_2)^2-\sigma_1^2}{(R-\sigma_2)^2-\sigma_1^2}\right)^{\frac{1}{4}}\left(\frac{R^2-(\sigma_1-\sigma_2)^2}{R^2-(\sigma_1+\sigma_2)^2}\right)^{\frac{9a^2}{4(3a^2+4)}}&=&1.
\label{consat}
\eeqs
In absence of the dilaton, when $a=0$ is it easy to see that one recovers the equilibrium condition for the Black Saturn in the Einstein-Maxwell theory, found previously in \cite{Chng:2008sr}. We performed a numerical analysis of this equation searching for various values of the parameters describing non-extremal conﬁgurations. Although a systematic analysis of this issue is beyond the purpose of this paper, we were so far unable to find reasonable values of the parameters to describe dilatonic black saturns in equilibrium. Let us notice that one could satisfy this conditions for extremal objects, however the price to pay is that the horizon of the black rings becomes a naked singularity.

It is also instructive to compare our general Black Saturn solution to the dilatonic solution that can be constructed starting from a static vacuum Black Saturn.\footnote{See for instance the solution with $c=0$ in equation ($39$) in \cite{Stelea:2011fj}, which was based on the general Harisson transformation in \cite{Kleihaus:2009ff}.} In the second case the generated charges $Q_i$ of the two black holes are proportional to their masses $Q_i=\delta M_i$. However, unlike in our more general solution, the coupling constant $a$ of the dilaton to the electromagnetic field has no influence on the presence of the conical singularities in that particular solution. As a consistency check of our more general easy to check that if the charges are proportional to the masses this happens in our more general solution as well. To see this, let us notice that for $Q_i=\delta M_i$ one obtains:
\beqs
\sigma_i&=&\sqrt{1-\delta^2}M_i,~~~~\mu=0,~~~~\nu=R^2-(M_1^2+M_2^2)(1+\delta^2),~~~k=M_1M_2(1-\delta^2).
\eeqs
Replacing these values in (\ref{consat}) one obtains:
\beqs
k_3&=&\left(\frac{(R+M_2\sqrt{1-\delta^2})^2-M_1^2(1-\delta^2)^2}{(R-M_2\sqrt{1-\delta^2})^2-M_1^2(1-\delta^2)^2}\right)^{\frac{1}{4}}\left(\frac{R^2-(M_1-M_2)^2(1+\delta^2)}{R^2-(M_1+M_2)^2(1+\delta^2)}\right)^{\frac{3}{4}},
\eeqs
which shows that indeed the dilaton coupling constant makes no appearance in the conical singularity condition.

\subsection{Some properties of the new solutions}

By construction, our solutions are asymptotically flat, as can be also be seen from their rod structure. Then the total mass and the total charge can be computed in the asymptotic region, which is reached by first performing the coordinate transformations:
\beqs
\rho&=&\frac{r^2}{2}\sin 2\theta,~~~~~z=\frac{r^2}{2}\cos 2\theta
\eeqs
then taking the $r\ra\infty$ limit. The easiest mode to compute the conserved charges such as the total mass ${\cal M}$, total charge ${\cal Q}$ and the dilaton charge $\Sigma$ is by using the asymptotic behavior of the metric, electromagnetic potential and the dilaton function:
\beqs
g_{tt}&=&-1+\frac{8{\cal M}}{3\pi}\frac{1}{r^2}+... ,\nonumber\\
A_t&=&\frac{4{\cal Q}}{\pi}\frac{1}{r^2}+... ,\nonumber\\
\phi&=&\frac{4\Sigma}{\pi}\frac{1}{r^2}+... ,
\eeqs
One obtains the following conserved charges:
\beqs
{\cal M}&=&\frac{3\pi}{4}\frac{1}{3a^2+4}\big[8(M_1+M_2)+3a^2(\sigma_1+\sigma_2)\big],\nonumber\\
{\cal Q}&=&\frac{\pi}{2}\sqrt{\frac{12}{3a^2+4}}(Q_1+Q_2),\nonumber\\
\Sigma&=&6\pi\frac{a}{3a^2+4}\big[M_1+M_2-2(\sigma_1+\sigma_2)\big].
\eeqs
The total charge receive contributions ${\cal Q}_i=\frac{\pi}{2}\sqrt{\frac{12}{3a^2+4}}Q_i$ from each black hole. Note that if the coupling constant $a=0$ then these expressions reduce to those previously found in literature. One can also compute the electric potential for each black hole horizon $\Phi^i_H=-A_t|_H$:
\beqs
\Phi^i_H&=&\sqrt{\frac{12}{3a^2+4}}\left(\frac{M_i-\sigma_i}{Q_i}\right).
\eeqs

Finally, for each black hole one can evaluate its horizon area:
\beqs
Area^i_H&=&8\pi^2\sigma_i(\rho f^{-1}e^{\lambda})^{\frac{3}{3a^2+4}}(\rho^{\frac{1}{2}}e^{2\tau-2H})^{\frac{1}{2}}(\rho^{\frac{1}{2}}e^{2\gamma-2h})^{\frac{9a^2}{2(3a^2+4)}}\equiv \frac{8\pi^2\sigma_i}{k^i_H},
\eeqs
where $k^i_H$ represent the surface gravities for each horizons, as calculated in the rod structure for our solutions. Then the entropy of each horizon can be easily computed as $S_i=Area^i_H/4$. The Hawking temperatures for each black hole can also be expressed in terms of the surface gravity using the usual formula $T_i=\frac{k^i_H}{2\pi}$ and one recovers the simple relation:
\beqs
S_iT_i&=&\pi\sigma_i.
\eeqs
One is now ready to verify that the Smarr relation is satisfied, as expected:
\beqs
\frac{2}{3}{\cal M}&=&S_1T_1+S_2T_2+\frac{2}{3}(\Phi^i_H{\cal Q}_1+\Phi^i_H{\cal Q}_2).
\eeqs

\section{Conclusions}

In the present paper we presented a new solution generating technique, which is a generalization of the one previously presented in \cite{Chng:2008sr}. This new method leads to new and more general solutions of the EMD theory in five dimensions. Unlike the previous known solution generating technique from \cite{Chng:2008sr}, which contained only one arbitrary harmonic function, in the present method the final solution is defined up to two arbitrary harmonic functions. As an illustrative example, we presented in some detail the derivation of the dilatonic Reissner-Nordstr\"{o}m solution in five dimensions. Then, using the four dimensional double-Reissner-Nordstr\"{o}m solution in the parameterization given by  Manko in \cite{Manko:2007hi}, \cite{Manko:2008gb}, we were be able to generate the general dilatonic charged double-black hole solutions in five dimensions. In particular, we discussed the case of two spherical black holes  and the proper dilatonic generalization of the charged Black Saturn solution. We investigated the effect of the dilaton field in the balance of forces in between the black holes. So far our numerical investigations failed to find reasonable values of the parameters for which the regularity conditions are satisfied and the conical singularities are eliminated. However, physically, the presence of these unavoidable conical singularities is to be expected since our solutions are static and, therefore, the conical singularities signal the presence of some other forces needed to balance the gravitational and electromagnetic forces in between the black holes. Finally, we discussed some of the properties of the new solutions: we computed the conserved charges and the physical quantities on the horizon and we verified that the Smarr relation is satisfied, as expected.

As avenues for further work, an interesting extension of the present work will be in the context of black holes in backgrounds with Kaluza-Klein asymptotics. Work on these matters is in progress and it will be presented elsewhere.

\vspace{10pt}

{\Large Acknowledgements}

The work of C. S. was financially supported by UEFISCDI through the PN-III-P4-ID-PCE-2016-0131 program.

\end{document}